\documentclass[journal]{IEEEtran}
\usepackage[usenames,dvipsnames]{color}
\let\labelindent\relax
\usepackage{enumitem}
\usepackage{latexsym,psfig}
\usepackage{amsfonts}
\usepackage{amsmath,amssymb}
\usepackage{amsmath,epsfig}
\RequirePackage{color}
\usepackage{amssymb}
\usepackage{amsthm}
\usepackage{amsfonts}
\usepackage{mathrsfs}
\usepackage{algorithm}
\usepackage{algpseudocode}
\usepackage{times,lastpage,bm,xspace,subfigure,booktabs,bbm}
\usepackage{bibunits}
\usepackage{multirow}
\usepackage{url}

\def\RIP{\text{RIP}}

\renewcommand{\vec}[1]{{#1}}
\newcommand{\mat}[1]{{\uppercase{#1}}}
\newcommand{\oper}[1]{{#1}}
\def\downsamp{\oper{D}}
\def\vect{{\textrm {vect}}}
\newcommand{\wh}[1]{{\widehat{#1}}}
\newcommand{\wt}[1]{{\widetilde{#1}}}

\def\FF{\mat{\mathcal{F}}}

\def\IF{\FF^{-1}}

\def\fstar{\vec{f}^{\star}}
\def\thetastar{\vec{\theta}^{\star}}

\def\f{\vec{f}}

\def\fest{\widehat{\f}}

\def\R{\mat{R}}

\def\y{\vec{y}}

\def\test{\widehat{\vec{\theta}}}

\def\hmura{h^{\rm MURA}}
\def\hrecon{\overline{h}\! \phantom{.}^{\rm MURA}}

\def\hus{h^{\rm US}}
\def\hbs{h^{\rm BS}}
\def\hup{h^{\rm UP}}
\def\hgs{h^{\rm GS}}
\def\Aus{A^{\rm US}}
\def\Abs{A^{\rm BS}}
\def\Aup{A^{\rm UP}}
\def\Ags{A^{\rm GS}}

\def \ie{{\em i.e., }}
\def \cf{{\em cf. }}
\def \eg{{\em e.g., }}
\def\ones{\ensuremath{\mathbf{1}}}
\def\reals{\ensuremath{\mathbb{R}}}
\def\expect{\ensuremath{\mathbb{E}}}
\def\prob{\mathbbm{P}}

\def\deq{\ensuremath{\triangleq}}
\def\grad{\ensuremath{\nabla}} 

\def\diag{\ensuremath{\mathop{\rm diag}}}

\def\argmin{\ensuremath{\mathop{\arg \min}}}
\DeclareMathOperator*{\st}{subject\ to}

\newtheorem{definition}{Definition}
\newtheorem{remark}{Remark}

\newtheorem{theorem}{Theorem}

\newenvironment{squishlist}
{\begin{list}{$\bullet$}
{ \setlength{\itemsep}{2pt}      \setlength{\parsep}{2pt}
  \setlength{\topsep}{0pt}       \setlength{\partopsep}{0pt}
  \setlength{\leftmargin}{1.5em} \setlength{\labelwidth}{1em}
  \setlength{\labelsep}{0.5em} } }
{\end{list}}

\begin{document}
%
\title{Spatio-temporal Compressed Sensing with Coded Apertures and Keyed Exposures}
%
%
%

\author{Zachary~T.~Harmany,~\IEEEmembership{Student Member,~IEEE,} 
Roummel~F.~Marcia,~\IEEEmembership{Member,~IEEE,}
        and~Rebecca~M.~Willett,~\IEEEmembership{Senior Member,~IEEE}
\thanks{This research is supported by DARPA Contract
    No. HR0011-04-C-0111, ONR Grant No. N00014-06-1-0610, DARPA
    Contract No. HR0011-06-C-0109, and NSF-DMS-08-11062.  
    Portions of this work were presented at the IEEE International Conference on Acoustic, Speech, and Signal Processing, March 2008, and at the SPIE Electronic Imaging Conference, January 2009,
    and SPIE Photonics Europe, April 2010.
}
\thanks{ Z.T.~Harmany and R.M.~Willett are with the Department of Electrical and Computer Engineering, Duke University, Durham, NC 27708 USA (e-mail: zth@duke.edu, willett@duke.edu).}
\thanks{R.F.~Marcia is with the School of Natural Sciences, University of California, Merced, CA 95343 USA (e-mail: rmarcia@ucmerced.edu).
}%
}

%
%

\markboth{}{}
%



\maketitle

\begin{abstract}
Optical systems which measure independent random projections of a scene according to compressed sensing (CS) theory face a myriad of practical challenges related to the size of the physical platform, photon efficiency, the need for high temporal resolution, and fast reconstruction in video settings. This paper describes a coded aperture and keyed exposure approach to compressive measurement in optical systems. The proposed projections satisfy the Restricted Isometry Property for sufficiently sparse scenes, and hence are compatible with theoretical guarantees on the video reconstruction quality. These concepts can be implemented in both space and time via either amplitude modulation or phase shifting, and this paper describes the relative merits of the two approaches in terms of theoretical performance, noise and hardware considerations, and experimental results. Fast numerical algorithms which account for the nonnegativity of the projections and temporal correlations in a video sequence are developed and applied to microscopy and short-wave infrared data.
\end{abstract}

\begin{IEEEkeywords}
Compressive sensing, coded apertures, convex optimization, sparsity, coded exposure, Toeplitz matrices
\end{IEEEkeywords}

%
\IEEEpeerreviewmaketitle

%
%
%
%

\section{Introduction}

\IEEEPARstart{T}{he} theory of compressed sensing (CS) suggests that we can collect high-resolution imagery with relatively few photodetectors or a small focal plane array (FPA) when the scene is sparse or compressible in some dictionary or basis and the measurements are chosen appropriately \cite{CS:candes1,CS:donoho}. There has been significant recent interest in building imaging systems in a variety of contexts to exploit CS (\cf \cite{Shankar:08, Coskun:10, Potter:10, Gu:08, SparseDNA, CS_DNAMicroarrays, LustigMRI, GPR, confocal, CSastronomy}). By designing optical sensors to collect measurements of a scene according to CS theory, we can use sophisticated computational methods to infer critical scene structure and content. One particularly famous example in optical imaging is the Single Pixel Camera \cite{riceCamera}, which collects individual projections of the scene sequentially. While these measurements are supported by the CS literature, there are several practical challenges associated with the tradeoff between temporal resolution physical system footprint. In this paper we describe an alternative approach to designing a low frame-rate {\em snapshot} CS camera which naturally parallelizes the compressive data acquisition. Our approach is based on two imaging techniques called coded apertures and keyed exposures, which we explain next.

Coded apertures \cite{ables, dicke} have a long history in low-light astronomical applications. Coded apertures were first developed to increase the amount of light hitting a detector in an optical system without sacrificing resolution (by, say, increasing the diameter of an opening in a pinhole camera). The basic idea is to use a mask, \ie an opaque rectangular plate with a specified pattern of openings, that allows significantly brighter observations with higher signal-to-noise ratio than those from conventional pinhole cameras \cite{ables, dicke}.  These masks encode the image before detection, inducing a more complicated point spread function than that associated with a pinhole aperture. The original scene is then recovered from the encoded observations in post-processing using an appropriate reconstruction algorithm which exploits the mask pattern.  These multiplexing techniques are particularly popular in astronomical \cite{caroli, skinner} and medical \cite{accorsi, gindi, meikle} applications because of their efficacy at wavelengths where lenses cannot be used, but recent work has also demonstrated their utility for collecting both high resolution images and object depth information simultaneously \cite{freeman}.

Keyed exposure (also called coded exposure \cite{codedExposure}, flutter shutter \cite{flutterShutter}, or coded strobing \cite{codedStrobing}) imaging was initially developed to facilitate motion deblurring in video using a relatively low frame rate. In some cases motion has been inferred from a single keyed exposure snapshot. The basic idea is that the camera sensor continuously collects light while the shutter is rapidly opened and closed; the shutter movement effectively modulates the motion blur point spread function, and with well-chosen shutter movement patterns it becomes possible to deblur moving objects. Similar effects can be achieved using a strobe light instead of moving a shutter. 

Despite the utility of the above methods in specific settings, they both face some limitations. The design of conventional coded apertures does not account for the inherent structure and compressibility of natural scenes, nor the potential for nonlinear reconstruction algorithms. Likewise, existing keyed exposure methods focus on direct (uncoded) measurements of the spatial content of the scene and have limited reconstruction capabilities, as we detail below. The Coded Aperture Keyed Exposure (CAKE) sensing paradigm we propose in this paper is designed to allow nonlinear high-resolution video reconstruction from relatively few measurements in more general settings.

\subsection{Problem Formulation}
\label{sec:probform}

We consider the problem of reconstructing an $N$-frame video sequence $\fstar$, where each frame is an $n_1 \times n_2$ two-dimensional image denoted $\fstar_t$. Using standard vector representation, we have that $\fstar_t \in \reals^n$ for $t = 1,\ldots,N$ where $n \deq n_1 n_2$ is the total number of pixels. As a result, the vector representation of the video sequence is $\fstar = (\fstar_1, \ldots, \fstar_N) \in \reals^{nN}$.

The observations $y$ of $\fstar$ are also acquired as a video sequence. We do not assume that the observations are acquired at the same rate at which we will ultimately reconstruct $\fstar$. In general, we assume $y$ is an $M$-frame video sequence, with each frame $y_k$ of size $m_1 \times m_2$. Similarly to $\fstar$, we have $y_k \in \reals^m$ for $k = 1,\ldots,M$, where $m \deq m_1 m_2$, therefore $y = (y_1, \ldots, y_M) \in \reals^{mM}$.

We observe $\fstar$ via a spatio-temporal \emph{sensing matrix} $A \in \reals^{mM \times nN}$ which linearly projects the spatio-temporal scene onto an $mM$-dimensional set of observations:
\begin{equation} \label{eq:obs}
	\vec{y} = A\fstar + w,
\end{equation}
where $w \in \reals^{mM}$ is noise associated with the physics of the sensor. 

CS optical imaging systems must be designed to meet several competing objectives:
\begin{squishlist}
\item The sensing matrix $A$ must satisfy some necessary criterion (such as the RIP, defined below) which provides theoretical guarantees on the accuracy with which we can estimate $\fstar$ from $y$.
\item The total number of measurements, $mM$, must be lower than the total number of pixels to be reconstructed, $nN$. This is achievable via compressive spatial acquisition ($m < n$), frame rate reduction ($M < N$), or simultaneous spatio-temporal compression.
\item The measurements $y$ must be {\em causally} related to the temporal scene $\fstar$, which restricts the structure of the projections $A$.
\item The optical measurements modeled by $A$ must be implementable in a way that results in a smaller, cheaper, more robust, or lower power system.
\item The sensing matrix structure must facilitate fast reconstruction algorithms.
\end{squishlist}
{\em This paper demonstrates that compressive Coded Aperture Keyed Exposure  systems achieve all these objectives.}

\subsection{Contributions} 

The primary contribution of this paper is the design and theoretical characterization of compressive Coded Aperture Keyed Exposure (CAKE) sensing. We explore amplitude modulating and phase shifting masks and describe theoretical and implementation aspects of both. We further describe how keyed exposure ideas can be used in conjunction with coded apertures to increase both the spatial and temporal resolution of video from relatively few measurements. We prove hitherto unknown theoretical properties of such systems and demonstrate their efficacy in several simulations. In addition, we discuss several important algorithmic aspects of our approach, including a mean-subtraction pre-processing step which allows us to sidestep challenging theoretical aspects associated with nonnegative sensing matrices (such as amplitude modulating coded apertures). This paper builds substantially upon earlier preliminary studies by the authors \cite{marciaICASSP,MarciaEUSIPCO,MarciaHW_SPIE2010} and related independent work by Romberg \cite{RombergToeplitz}.

\subsection{Organization of the Paper} 

The paper is organized as follows. Section~\ref{sec:arch} describes conventional coded aperture imaging techniques and Section~\ref{sec:codedexposure} describes keyed exposure techniques currently in the literature. We describe the compressive sensing problem and formulate it mathematically in Section~\ref{sec:CS}. In Section~\ref{sec:cca}, we show how CS theory can be used for constructing coded aperture masks that can easily be implemented for improving image reconstruction resolution in snapshot imaging; this includes theoretical results, a discussion of implementation details and tradeoffs in practical optical systems, and experimental results. We consider applications to video compressed sensing using the full CAKE paradigm in Section~\ref{sec:video}, including theory and experimental results.

\section{Background}

Prior to detailing our main contributions, we first review pertinent background material. This review touches upon the development of coded aperture imaging, coded exposure photography, and a brief review of salient concepts in compressed sensing theory.

\subsection{Coded Aperture Imaging} 
\label{sec:arch}

Seminal work in coded aperture imaging includes the development of masks based on Hadamard transform optics \cite{Sloane:76} and pseudorandom phase masks \cite{Ashok:07}. Modified Uniformly Redundant Arrays (MURAs) \cite{mura} are generally accepted as optimal mask patterns for coded aperture imaging.  These mask patterns (which we denote by $\hmura$) are binary, square patterns, whose \emph{grid size matches the spatial resolution of the photo-detector}. Each mask pattern is specifically designed to have a complementary pattern $\hrecon$ such that $\hmura * \hrecon$ is a single peak with flat side-lobes (\ie a Kronecker $\delta$ function).

In practice, the resolution of a detector array dictates the properties of the mask pattern and hence resolution at which $\fstar$ can be reconstructed. We model this effect as $\fstar$ being downsampled to the resolution of the detector array and then convolved with the mask pattern $\hmura$, which has the same resolution as the FPA and the downsampled $\fstar$, \ie
\begin{equation} \label{eq:mura}
	y = ( \downsamp_\text{int} \fstar) * \hmura + w,
\end{equation}
where $*$ denotes circular convolution, $w$ corresponds to noise associated with the physics of the sensor, and $\downsamp_\text{int} \fstar$ is the \emph{integration} downsampling of the scene, which consists of partitioning $\fstar$ into uniformly sized $d_1 \times d_2$ blocks, where $d_i \deq n_i/m_i$ for $i = 1,2$, and measuring the total intensity in each block.

Because of the construction of $\hmura$ and $\hrecon$, $\downsamp_\text{int}\fstar$
can be reconstructed using
\begin{equation*}
	\fest = y * \hrecon.
\end{equation*}
However, the resulting resolution is often lower than what is necessary to capture some of the desired details in the image. Clearly, the estimates from MURA reconstruction are limited by the spatial resolution of the photo-detector.  Thus, high resolution reconstructions cannot generally be obtained from low-resolution MURA-coded observations. 

It can be shown that this mask design and reconstruction result in minimal reconstruction errors \emph{at the FPA resolution} and \emph{subject to the constraint that linear, convolution-based reconstruction methods would be used}.  However, when the scene of interest is sparse or compressible, and nonlinear sparse reconstruction methods may be employed, then CS ideas can be used to design coded aperture which yield higher resolution images. Before describing the details of this, we briefly review two key relevant concepts from CS.

\subsection{Coded (Keyed) Exposure Imaging}
\label{sec:codedexposure}

Coded (or keyed) exposures  were developed recently in the computational photography community. Initial work in this area was focused on engineering the temporal component of a motion blur point spread function by rapidly opening and closing the shutter during a single exposure or a small number of exposures at a low frame rate \cite{codedExposure, flutterShutter}. That is,
\begin{equation*}
y = A^{\rm KE} \fstar + w = \sum_{i \in S} \fstar_{i} + w,
\end{equation*}
where the keyed exposure (KE) measurement matrix $A^{\rm KE}$ selects the subset of frames
during which the shutter is open.  We refer to this subset as the 
exposure code $S \subseteq \{1,\ldots,N\}$.

If an object is moving during image acquisition, then a static shutter would induce a typical motion blur, making the moving object difficult to resolve with standard deblurring methods. However, by ``fluttering'' the shutter during the exposure using carefully designed patterns, the induced motion blur can be made invertible and moving objects can be accurately reconstructed. Instead of a moving shutter, more recent work uses a strobe light to produce a similar effect \cite{codedStrobing}.

While this novel approach to video acquisition can produce very accurate deblurred images of moving objects, there is significant overhead associated with the reconstruction process. To see why, note that every object moving with a different velocity or trajectory will produce a different motion blur. This means that (a) any stationary background must be removed during preprocessing and (b) multiple moving objects must be separated and processed individually. 

More recently, it was shown that these challenges could be sidestepped when the video is temporally periodic (\eg consider a video of an electronic toothbrush spinning) \cite{codedStrobing}. The periodic assumption amounts to a sparse temporal Fourier transform of the video, and this approach, called coded strobing, is a compressive acquisition in the temporal domain. As a result, the authors were able to leverage ideas from compressed sensing to achieve high-quality video reconstruction.

The assumption of a periodic video makes it possible to apply much more general reconstruction algorithms that do not require background subtraction or separating different moving components. However, it is a very strong assumption, which places some limits in its applicability to real-world settings. The approach described in this paper has similar performance guarantees but operates on much more general video sequences.

\subsection{Compressed Sensing}
\label{sec:CS}

In this section we briefly define the Restricted Isometry Property (RIP) and explain its significance to reconstruction performance.  In subsequent sections, we demonstrate our primary theoretical contribution, which is to prove the RIP for compressive coded aperture and keyed exposure systems.

\begin{definition}[Restricted Isometry Property (RIP) \cite{RIP}]
A matrix $A$ satisfies the RIP of order $s$ if there exists a constant $\delta_s \in (0,1)$ for which
\begin{equation} \label{eq:RIP}
  (1 - \delta_{s}) \| f \|_2^2 \le \|A f \|_2^2 \le
  (1 + \delta_{s}) \| f \|_2^2.
\end{equation}
holds for all $s$-sparse $f \in \reals^n$. If this property holds, we say $A$ is $\RIP(s,\delta_{s}).$
\end{definition}
Matrices which satisfy the RIP  are called CS matrices; and when combined with sparse recovery algorithms, they are guaranteed to yield accurate estimates of the underlying function $\fstar$:
\begin{theorem}[Sparse Recovery with RIP Matrices \cite{candesTutorial,Candes:06c}]
\label{thm:recovery2}
Let $A$ be a matrix satisfying $\mbox{RIP}(2s,\delta_{2s})$ with $\delta_{2s} < \sqrt{2}-1$, and let $y = Af + w$ be a vector of noisy observations of any signal $f \in \reals^{n}$, where the $w$ is a noise or error term with $\|w\|_{2}\leq \epsilon$ for some $\epsilon > 0$. Let $f_s$ be the best $s$-sparse approximation of $f$; that is, $f_s$ is the approximation obtained by keeping the $s$ largest entries of $f$ and setting the others to zero. Then the estimate
\begin{equation} \label{eq:constrained}
\begin{aligned}
\fest \ = \ &\argmin_{f \in \reals^n} 	& &\|f\|_{1}\\
		 &\st						& &\|y-Af\|_{2} \leq \epsilon,
\end{aligned}
\end{equation}
obeys
\begin{equation*}
\| f - \fest \|_{2} \leq C_{1,s} \epsilon + C_{1,s} \frac{\|f-f_{s}\|_{1}}{\sqrt{s}},
\end{equation*}
where $C_{1,s}$ and $C_{2,s}$ are constants which depend on $s$ but not on $n$ or $m$.
\end{theorem}
\noindent Note that the reconstruction \eqref{eq:constrained} in Theorem~\ref{thm:recovery2} is equivalent to
\begin{equation} \label{eq:l1regularized}
\fest = \argmin_{f \in \reals^n} \tfrac{1}{2}\|y -Af\|_2^2 + \tau \| f\|_1 
\end{equation}
where $\tau>0$, which depends on $\epsilon$, can be viewed as a regularization parameter.

\section{Compressive Coded Apertures for Snapshot Imaging}
\label{sec:cca} 

This section first considers a snapshot acquisition model. Our goal is to recover a static high-resolution scene from a single image where all pixels are collected simultaneously. In terms of the notation in Sec.~\ref{sec:probform}, we have $N=M=1$, so that $A$ is of size $m \times n$. The sensing matrix for compressive coded aperture (CCA) systems can be modeled mathematically as
\begin{equation} \label{eq:cca}
	A \fstar = \downsamp(\fstar * h);
\end{equation}
where $h$ is a coding mask, and $\downsamp$ is a subsampling operator (detailed below). Here the coding mask, $h$, is at the size and resolution at which $\fstar$ will be reconstructed; this is in contradistinction to the MURA system, in which $\hmura$ is at the size and resolution of the FPA. Thus in \eqref{eq:cca}, we model the measurements as the scene being convolved with the coded mask and \emph{then} downsampled. 

Using a similar model, Romberg \cite{RombergToeplitz} conducted related work concurrent and independent of our initial investigations \cite{marciaICASSP}. We will summarize the key features of this model and compare with our approach in Sec.~\ref{sec:psmcca}. While these models share common elements, we will see in Sec.~\ref{sec:hardware} that there are important tradeoffs associated with the theory and implementation of each strategy.

Recent work by Bajwa et al.~\cite{waheed, haupttoeplitz}, showed that random circulant matrices (and Toeplitz matrices, in general) are sufficient to recover sparse $\f^{\star}$ from $\y$ with high probability.  In particular, they showed that a Toeplitz matrix whose first row contains elements drawn independently from a Gaussian distribution are $\RIP(s,\delta_s)$ when $m\geq C s^2 \log(n)$ for some constant $C$. 
Here, we extend these results to pseudo-circulant matrices and use them to motivate our mask design. Our model differs from that in \cite{RombergToeplitz} in that we consider a different generative model for the coded aperture mask, as well as a different subsampling strategy:
\begin{squishlist}
\item The elements of the coding mask $h$ are generated iid according to a particular generating distribution (\eg an appropriately scaled Rademacher, uniform, or Gaussian distribution). 
\item Our analysis allows for deterministic downsampling in which we collect one sample per nonoverlapping block. 
\end{squishlist}
In our notation, we distinguish mask patterns in this manner using the abbreviations BS, US, and GS, where the first character denotes the distribution used to generate the mask, \ie binary, uniform, or Gaussian, and the second is a reminder that these masks are generated directly in the spatial domain. 

\subsection{CCAs Generated in the Spatial Domain}
\label{sec:ammcca}

\begin{figure}
\begin{center}
\includegraphics[width=80mm]{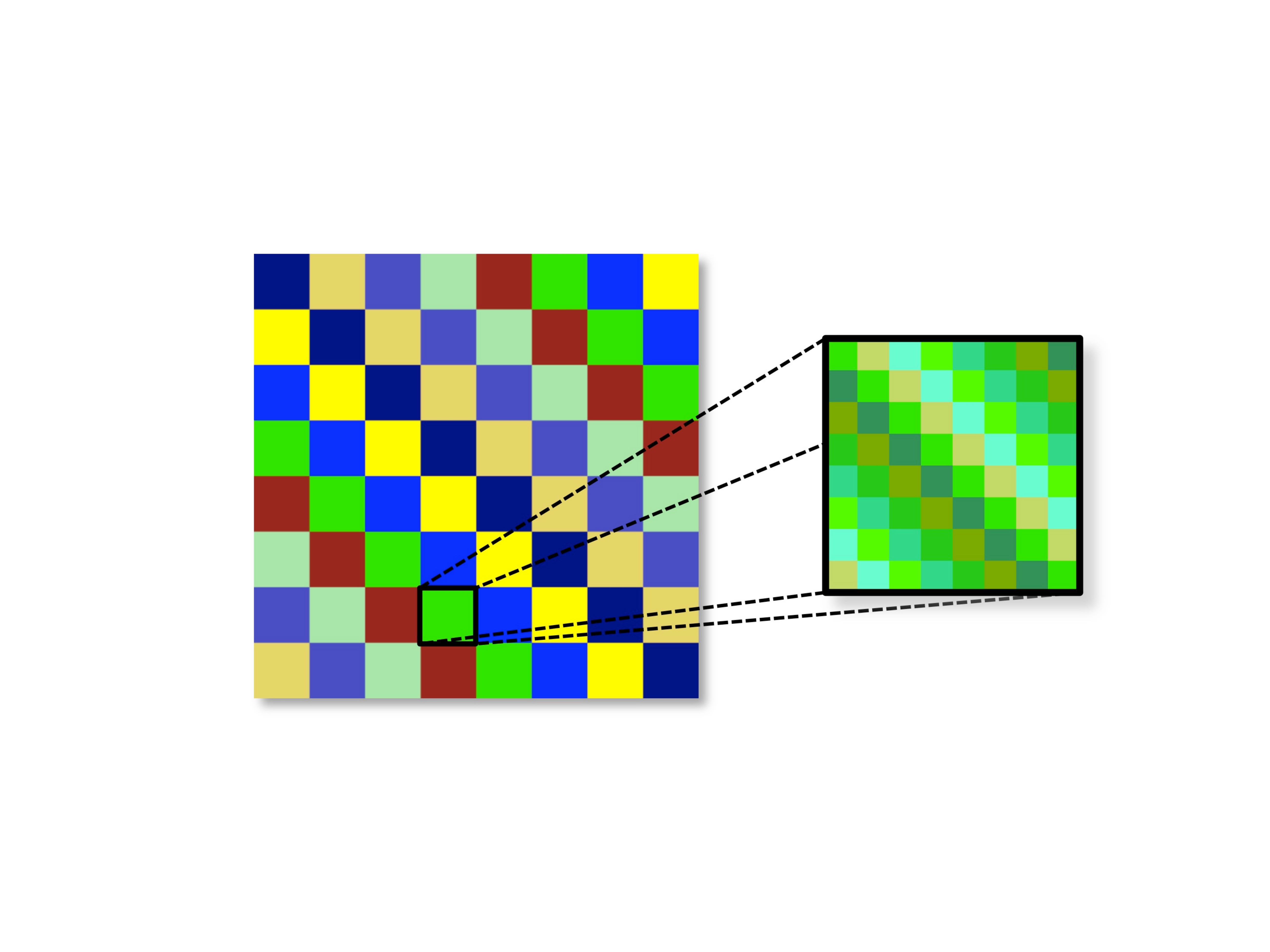}
\caption{The $n \times n$ matrix $\IF \diag(\FF h) \FF$ is block-circulant with $n_2$ blocks in each row and column. Each block is $n_1 \times n_1$ and is circulant.}
\label{fig:circulant} 
\end{center}
\end{figure}

The two-dimensional convolution of $h$ with an image $\fstar$ as in (\ref{eq:cca}) can be represented as the application of the Fourier transform to $\fstar$ and $h$, followed by element-wise multiplication and application of the inverse Fourier transform. In matrix notation, this series of linear operations can be expressed as
\begin{equation} \label{eq:R}
	\vect(\fstar * h)  = \IF \diag(\FF h) \FF \fstar = R \fstar,
\end{equation}
where $\vect(f)$ is a vectorized representation of an image $f$, $\FF$ is the two-dimensional Fourier transform matrix, and $\diag(\FF h)$ is a diagonal matrix whose elements correspond to the transfer function, which is the Fourier transform of $h$. The matrix product $\R = \IF \diag(\FF h) \FF \in \reals^{n \times n}$ is block-circulant and each block is in turn circulant. In matrix notation, $\R$ is consists of $n_2 \times n_2$ blocks,
\begin{equation}
	\R= 
	\begin{bmatrix}
		\R_{1} 		& \R_{n_2}		& \cdots & \R_3 		& \R_2 \\
		\R_{2} 		& \R_1 			& \cdots & \R_4 		& \R_3 \\
		\vdots 		& \vdots 		& \ddots & \ddots 	& \vdots \\
		\R_{n_2} 	& \R_{n_2-1} 	& \cdots & \R_2  	& \R_1
	\end{bmatrix},
\label{eq:blockcirc}
\end{equation}
where each $\R_j \in \reals^{n_1 \times n_1}$ is circulant; \ie
$\R_j$ is of the form
\begin{equation*}
	\R_j =
	\begin{bmatrix}
		r_{j,1} 		& r_{j,n_1} 		& \cdots & r_{j,3}	& r_{j,2} \\
		r_{j,2} 		& r_{j,1} 		& \cdots & r_{j,4} 	& r_{j,3} \\
		\vdots 		& \vdots 		& \ddots & \ddots 	& \vdots  \\
		r_{j,n_1} 	& r_{j,n_1-1} 	& \cdots & r_{j,2} 	& r_{j,1} 
	\end{bmatrix},
\end{equation*}
(see Fig.~\ref{fig:circulant}).
This block-circulant with circulant-block (BCCB) structure of $\R$ is a direct result of the fact that the $k$-point one-dimensional Fourier transform $F_k$ diagonalizes any $k \times k$ circulant matrix (such as $\R_j$ with $k=n_1$) and so ${\FF} \equiv F_{n_2} \otimes F_{n_1}$ diagonalizes block-circulant matrices (such as $\R$). Here $\otimes$ denotes the Kronecker matrix product.

We now examine the generation of a compressed sensing matrix from the convolution $R$ by restricting the number of measurements collected of the vector $R\fstar$. Here we simply subsample the vector $R \fstar$ by applying a pointwise subsampling matrix $\downsamp_\text{sub}$. The operation of applying $\downsamp_\text{sub}$ consists of retaining only one measurement per uniformly sized $d_1 \times d_2$ block, \ie we subsample by $d_1$ in the first coordinate and $d_2$ in the second coordinate so that the result is an $m_1 \times m_2$ image with $m_i = n_i/d_i, i = 1,2$. 
In matrix form $\downsamp_\text{sub}$ can be thought of as retaining a certain number of rows of the identity matrix. 
Because of the structure and deterministic nature of this type of downsampling, it is often more straightforward to realize in practical imaging hardware (see Sec.~\ref{sec:hardware}).

The resulting projection matrix $A$ is then given by
\begin{equation*}
A = \downsamp_\text{sub} \R = \downsamp_\text{sub} \IF \diag(\FF h) \FF.
\end{equation*}
We now show that if the elements of $h$ are drawn from an appropriate probability
distribution, then $A$ will be $\RIP(s,\delta_s)$ with high probability.

\begin{theorem}[Spatial-Domain CCA Sensing]
\label{thm:BCCB} Let $\hbs$ be a mask with entries generated i.i.d.\ according to the scaled Rademacher distribution 
\begin{equation} \label{eq:rademacher}
\hbs_{k_1,k_2} = \begin{cases} \phantom{-}\sqrt{d/n} & \text{ with probability } 1/2, \\
                                         -\sqrt{d/n} & \text{ with probability } 1/2, \end{cases}
\end{equation}
for $k_1 = 1, \ldots, n_1, k_2 = 1, \ldots, n_2$.
Let $\Abs = \downsamp_\text{sub} R$ be an $m \times n$ matrix, 
where $R \in \reals^{n \times n}$ is the BCCB matrix generated from $\hbs$, and $\downsamp_\text{sub} \in \reals^{m \times n}$ is the pointwise subsampling matrix.
Then, there exists constants $c_1, c_2 \ge 0$ depending on $\delta_{s}$ such that for any 
\begin{equation}
m \ge c_1 s^2 \log(n),
\end{equation} $\Abs$ is $\RIP(s,\delta_s)$ with probability exceeding 
\begin{equation} \label{eq:RIP_prob}
	1 - 2n^2 e^{-c_2 m/s^2}.
\end{equation}
\end{theorem}
\begin{remark}
 This binary-valued distribution was selected to model coded apertures with two states -- open and closed -- per mask element.  We discuss the issue of implementing these masks in Section~\ref{sec:maskgeneration}.
 \end{remark}
\begin{remark}
 It is straightforward to extend the result to other mask generating distributions, such as uniform and Gaussian, which we denote $\hus$ and $\hgs$, with associated sensing matrices $\Aus$ and $\Ags$.
 \end{remark}
We present the proof of Theorem~\ref{thm:BCCB} in Appendix~\ref{sec:proof}.

For successful recovery, the coded aperture masks are designed to be satisfy $\RIP(2s,\delta_{2s})$ as described in \eqref{eq:RIP} with high probability when $m \ge c_1 s^2 \log(n)$. Note that this is somewhat weaker than what can be achieved by a purely unstructured i.i.d. randomly generated sensing matrix which satisfies \eqref{eq:RIP} with high probability when $m \geq \tilde{c}_1 s \log(n/s)$ for some constant $\tilde{c}_1 > 0$. Intuition may lead one to believe the extra factor of $s$ is due to the fact that the $m$ projections sensed using the amplitude modulated mask framework exhibit dependencies. However, in many settings this theoretical disadvantage is offset by advantageous practical implementations.

\subsubsection*{Sparse Gradients Scenes}
\label{sec:grad}

The above theory is applicable to reconstructing $\fstar$ when $\fstar$ is sparse in the pixel basis; similar results hold when the {\em gradient} of $\fstar$ is sparse. For simplicity of notation we will show then in a 1d setting; the extension to 2d is straightforward. Let $\nabla f$ denote the first-order gradient of $f$, so that
\begin{equation}
\nabla \deq 
\begin{bmatrix}
\ \ 1 &  \ \ 0 & \ \ 0 & \cdots & \  0 \ \\
-1 & \ \ 1 & \ \ 0 & \ddots &  \  \vdots \ \\
\ \ 0 & -1 & \ \ 1 & \ddots & \  0  \ \\
\ \ \vdots & \ddots & \ddots & \ddots & \  0 \ \\
\ \ 0 & \cdots & \ \ 0 & -1 & \  1 \phantom{\vdots}
 \end{bmatrix}.
 \label{eq:grad}
 \end{equation}

As noted in \cite{waheed}, $\fstar = \nabla^{-1} \theta$, where $\theta$ is the sparse gradient image. Thus if we sense $y = A\nabla \fstar + w = A\nabla \nabla^{-1} \theta + w = A \theta + w$, we can expect to recover $\theta$ using sparse reconstruction methods. 

\subsection{CCAs Generated in the Fourier Domain}
\label{sec:psmcca}

%

The random convolution in \cite{RombergToeplitz} follows the same structure as that in Sec.~\ref{sec:ammcca}.  However, in that work the convolution is generated randomly in the frequency domain. More specifically, the entries of the transfer function correspond to random \emph{phase shifts} (with some constraints to keep the resulting observation matrix real-valued). We denote the resulting convolution kernel as $\hup$, where ``UP'' refers to ``uniform phase''. For simplicity of presentation, we describe the generating distribution for a one-dimensional convolution for even-length masks. In particular, let $\sigma = F \hup$, and generate $\sigma$ such that for $k = 1,\ldots,n$,
\begin{equation}
\sigma_k = \begin{cases} \pm 1 \text{ with equal probability}	& \text{if $k=1$}, \\
						e^{i \phi} \text{ with $\phi \sim \mathcal{U}(0, 2\pi)$} & \text{if $2 \le k \le n/2$}, \\
						\pm 1 \text{ with equal probability}	& \text{if $k=n/2+1$} \\
						\sigma_{n-k+2}^* & \text{if $n/2 + 2 \le k \le n$}.
			\end{cases}
\label{eq:psmdist}
\end{equation}
Here $\mathcal{U}(0, 2\pi)$ denotes the uniform distribution over $[0, 2 \pi]$. The real-valued convolution kernel is then given by $\hup = F^{-1} \sigma$. This result can be extended easily for two-dimensional convolutions. 

To form a compressed sensing matrix from this random convolution, \cite{RombergToeplitz} considers two different downsampling strategies: sampling at random locations, and random demodulation. The first method entails selecting a random subset $\Omega \subset \{1,\ldots,n\}$ of indices. 
We form a downsampling matrix $\downsamp_\Omega$ by retaining only the rows of the identity matrix indexed by $\Omega$, hence the resulting measurement matrix is given by
\begin{equation}
\Aup= \downsamp_\Omega \IF \diag(\sigma) \FF,
\label{eq:psmsubsample}
\end{equation}
where $\sigma$ is generated according to the two-dimensional analog of \eqref{eq:psmdist}. The random demodulation method multiplies the result of the random convolution by a random sign sequence $s \in \{-1,1\}^n$, such that $s_i = \pm 1$ with equal probability for all $i = 1, \ldots, n$, then performs an integration downsampling of the result. Therefore in this case the measurement matrix is
\begin{equation}
\Aup = \downsamp_\text{int} \diag(s) \IF \diag(\sigma) \FF.
\label{eq:psmdemodulate}
\end{equation}
It can be shown that both of these strategies yield RIP-satisfying matrices with high probability. 

\begin{theorem}[Fourier-Domain CCA Sensing]
Let $\Aup$ be an $m \times n$ sensing matrix resulting from random convolution with phase shifts followed by a downsampling strategy, and let $W$ denote an arbitrary orthonormal basis.
If the downsampling is random subsampling as in \eqref{eq:psmsubsample}, then there exists constant $c_3>0$ such that with probability exceeding $1 - O(n^{-1})$, for
\begin{equation*}
m \ge c_3 \delta^{-2} \min(s \log^6 n, s^2 \log^2 n),
\end{equation*}
$\Aup W$ satisfies $\RIP(2s,\delta_{2s})$ with $\delta_{2s} \le \delta$.
If the downsampling is random demodulation as in \eqref{eq:psmdemodulate}, then there exists constant $c_4>0$ such that with probability exceeding $1-O(n^{-1})$, for
\begin{equation*}
m \ge c_4 \delta^{-2} \min(s \log^6 n, s^2 \log n),
\end{equation*}
$\Aup W$ satisfies $\RIP(2s,\delta_{2s})$ with $\delta_{2s} \le \delta$.
 \end{theorem}

These theoretical results are stronger than those in Theorem~\ref{thm:BCCB}, 
especially since they allow sparsity in arbitrary orthonormal bases. However, there are important
differences between the observation models which have a significant impact on the feasibility and ease of hardware implementation. We
elaborate on this in Section~\ref{sec:hardware}. Nevertheless, the theory developed in \cite{RombergToeplitz} lends important theoretical
support to the general concept of compressive coded apertures.


\subsection{Hardware and Practical Implementation Considerations}
\label{sec:hardware}
	
In this section we describe how we shift from modeling the coded aperture masks in a way that is compatible with compressed sensing theory, to a model that describes their actual implementation in an optical system. Our analysis does not account for the bandwidth of the lenses; in particular, we implicitly assume that the bandwidth of the lenses is high enough that band-limitation effects are negligible at the resolution of interest. In all of the hardware settings described below, precise alignment of the optical components (\eg the mask and the focal plane array) is critical to the performance of the proposed system. Often a high-resolution FPA is helpful for alignment and calibration.

In this paper we focus on {\em incoherent} light settings (consistent with many applications in astronomy, microscopy, and infrared imaging). In this case, the coded aperture must be real-valued and flux-preserving (\ie the light intensity hitting the detection cannot exceed the light intensity of the source). In this section, we consider the following apertures:
\begin{squishlist}
\item{\bf Binary Spatial Mask:} $h \in \{0,1/n\}^{n_{1} \times n_{2}}$, drawn with equal probability,
\item{\bf Uniform Spatial Mask:} $h  \in [0,1/n]^{n_{1} \times n_{2}}$, where each element is drawn independently from a uniform distribution,  or
\item{\bf Uniform Phase Mask:} $h = |\FF p|^{2}$ for some $p$, where $p$ corresponds to a phase-shifting mask in an incoherent light setting.
\end{squishlist}

\subsubsection{Amplitude Modulation Masks}
\label{sec:maskgeneration}

In a conventional lensless coded aperture imaging setup, the point spread function associated with the aperture is the mask pattern $h$ itself. 
To shift a RIP-satisfying aperture as in Theorem \ref{thm:BCCB} to an implementable aperture, one simply needs to apply an affine transform to $h$ mapping $[-\sqrt{d/n},\sqrt{d/n}]$ to $[0, 1/n]$. 
This transform ensures that the resulting mask pattern is nonnegative and flux-preserving.


These amplitude modulating masks may be implemented using a spatial light modulator (SLM) or placing chrome on quartz. The SLMs may be preferable in video settings where the underlying scene contains motion and using a different mask pattern at each time step boosts performance. However, in order for the proposed approach to work, the mask or SLM used must be higher resolution than the FPA. Currently, very high resolution SLMs are still in development. Chrome on quartz masks can be made with higher resolution than many SLMs, but cannot be changed on the fly unless we mount a small number of fixed masks on a rotating wheel or translating stage. The uniform amplitude modulation masks in particular could be constructed using a high-resolution halftoning procedure, which is easiest to implement at the necessary resolution with chrome on quartz.

Both Robucci {\it et al.} \cite{Robucci:10} and Majidzadeh {\it et al.} \cite{Jacques:10} have proposed performing the analog, random convolution step in complementary, metal-oxide-semiconductor (CMOS) electronics.  A clear advantage to this architecture is that the additional optics required for spatial light modulation are removed in favor of additional circuitry, immediately reducting imager size.  

\subsubsection{Phase Shift Masks}

Phase shifting masks for coded aperture imaging have been implemented recently using a phase screen \cite{phaseCode}. This approach allows one to account for diffraction in the optical design. However, depending on the precise optical architecture, phase shift masks may be much less photon-efficient than amplitude modulation masks. Additionally, the mask generation distribution described in Eq.~\eqref{eq:psmdist} will result in negative entries for the corresponding PSF $\hup = \FF^{-1}\sigma$. To compensate for this, the phase mask must be mean-shifted to make all entries nonnegative, so that we are actually implementing
\begin{equation}
\hup_+ = c(\hup - \min(\hup)\ones),
\end{equation}
with the constant $c$ selected so that the implementable PSF $\hup_+$ is flux-preserving and $\ones$
is a matrix of ones and of the same dimension as $\hup$. 

\subsubsection{Implementable Masks for Scenes with Sparse Gradients}
\label{sec:diffs}
As described in Section~\ref{sec:grad}, theoretical results for scenes with sparse gradients hold for the sparse difference operator $\nabla$ defined in \eqref{eq:grad}. The problem with this approach is that $\nabla$ is invertible but not circulant -- and hence the sensing matrix $A\nabla$ cannot be implemented with a coded aperture system. We address this by noting that if the upper right element of $\nabla$ were $-1$ instead of $0$, then the resulting sensing matrix, denoted $\wt{\nabla}$, could be implemented physically and is a close approximation to the theoretically supported $\nabla$. In short, the theoretically supported solution is to set
\begin{align*}
	y &= A \nabla \fstar + w \\
	\test &= \argmin_{\theta} \tfrac{1}{2}\| y - A \theta \|_{2}^{2} + \tau\|\theta\|_1 \\
	\fest &= \nabla^{-1}\test\\
	\intertext{while an implementable close approximation is to set $A^\text{G} \deq A\wt{\nabla}$ and}
	y &= A^\text{G} \fstar + w \\
	\fest &= \argmin_{f} \tfrac{1}{2}\| y - A^\text{G} f\|_{2}^{2} + \tau \| f \|_{\rm TV},
\end{align*} 
where $\| \cdot \|_{\rm TV}$ is the total variation seminorm which causes $\fest$ to have sparse gradients. In our numerical experiments we compare the performance of TV regularization to sparsity penalization for the task of reconstructing static scenes. 

\subsubsection{Downsampling Implementation}

In developing RIP-satisfying AMM coded apertures, Theorem \ref{thm:BCCB} assumes the subsampling operation selects one measurement per $d_1 \times d_2$ block. From an implementation standpoint, this operation effectively discards a large portion ($(d-1)/d$) of the available light, which would result in much lower signal-to-noise ratios at the detector. A more pragmatic approach is to use larger detector elements that essentially sum the intensity over each $d_1 \times d_2$ block, making a better use of the available light. We call this operation \emph{integration downsampling} to distinguish it from subsampling. The drawback to this approach is that we lose many of the desirable features of the system in terms of the RIP. Integration downsampling causes a large coherence between neighboring columns of the resulting sensing matrix $A$. 
An intermediate approach would randomly sum a fraction of the elements in each size $d$ block, which increases the signal-to-noise ratio versus subsampling, but yields smaller expected coherence. This approach is motivated by the random demodulation proposed in \cite{RombergToeplitz} and described in Sec~\ref{sec:psmcca}, whereby the signal is multiplied by a random sequence of signs $\{-1, +1\}$, then block-wise averaged. The pseudo-random summation proposed here can be thought of as an optically realizable instantiation of the same idea where we multiply by a random binary $\{0, 1\}$ sequence.  We explore the effects  of these choices in our numerical results section.
\subsubsection{Noise and Quantization}
\label{sec:noise}
While CS is particularly useful when the FPA needs to be kept compact,
it should be noted that CS is more sensitive to measurement errors and
noise than more direct imaging techniques.  The experiments
conducted in this paper simulated very high signal-to-noise ratio
(SNR) settings and showed that CS methods can help resolve high
resolution features in images. However, in low SNR settings CS
reconstructions can exhibit significant artifacts that may even cause
more distortion than the low-resolution effects associated with
conventional coded aperture techniques such as MURA.

Similar observations are made in \cite{CSvsConventional}, which
presents a direct comparison of the noise robustness of CS in contrast
to conventional imaging techniques both in terms of bounds on how
reconstruction error decays with the number of measurements and in a
simulation setup; the authors conclude that for most real-world images, CS
yields the biggest gains in high signal-to-noise ratio (SNR)
settings. Related theoretical work in \cite{raginskyWillettPCS_TSP}
show that in the presence of low SNR photon noise, theoretical error
bounds can be large, and thus the expected performance of CS may be
limited unless the number of available photons to sense is
sufficiently high. These considerations play an important role in choosing the type of downsampling to implement.

Similar issues arise when considering the
bit-depth of focal plane arrays, which corresponds to measurement
quantization errors. Future efforts in designing optical CS systems
must carefully consider the amount of noise anticipated in the
measurements to find the optimal tradeoff between the focal plane
array size and image quality. 

\subsection{Reconstruction}
\label{sec:algo}

To solve the CS minimization problem \eqref{eq:l1regularized}, we use well-established
gradient-based optimization methods.  They are particularly suitable in our setting because 
the block-circulant structure of $A$ described in the previous section
allows for very fast matrix-vector products that are critical to the speed of their performance.
However, our observation matix $A$ is not zero mean and, therefore, negatively impacts the 
performance of these CS-based reconstruction algorithms.  In this section, we describe how
we take full advantage of the structure of $A$ and address how
we mitigate the negative effects of $A$ having nonzero-mean.

\subsubsection{Sparsity-promoting Methods}

We note that most popular and effective methods for
performing the above reconstruction are iterative algorithms with
repeated applications of the operators $A$ and $(A)^T$ to
scene estimates and residuals \cite{sparsa,gpsr}. In most CS settings, computing these
matrix-vector multiplications is a large computational
burden. However, because of the circulant structure of $A$, this
computation is very efficient using fast Fourier transform algorithms. 
In particular, we
compared the computation time for reconstructing a $256 \times 256$
scene using $128 \times 128$ CCA measurements with the time required
to reconstruct the same scene using the same number of measurements
computing using a dense random projection matrix; our reconstruction was 250 times
faster because of our exploitation of the Toeplitz structure in
$A$.

\newlength{\smallreconfigwidth}
\setlength{\smallreconfigwidth}{.945in}
\newlength{\biglabelwidth}
\setlength{\biglabelwidth}{.38\smallreconfigwidth}

\begin{figure}[t!]
\centering \footnotesize
\begin{tabular}{@{}c@{~}c@{~}c@{~}c@{~}}
&
\multicolumn{3}{c}{
\begin{tabular}{c@{~}c@{~}}
\normalsize Ground Truth  & \normalsize Conventional\\
\includegraphics[width=\smallreconfigwidth]{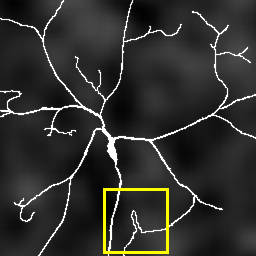}  & 
\includegraphics[width=\smallreconfigwidth]{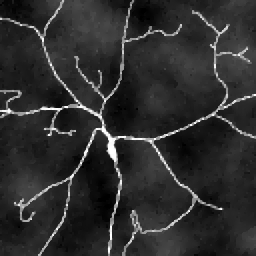} \\
\includegraphics[width=\smallreconfigwidth]{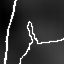}  & 
\includegraphics[width=\smallreconfigwidth]{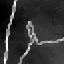} 
\end{tabular} \ \ } \vspace{.2cm}\\
& \multicolumn{3}{c}{\normalsize Different mask distributions}\\
& \normalsize Binary Spatial & \normalsize Uniform Spatial & \normalsize Uniform Phase\\
\multirow{1}{*}{\rotatebox{90}{\makebox[\biglabelwidth][c]{\normalsize Integration Downsampling}}}  &
\includegraphics[width=\smallreconfigwidth]{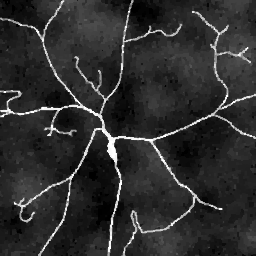} &
\includegraphics[width=\smallreconfigwidth]{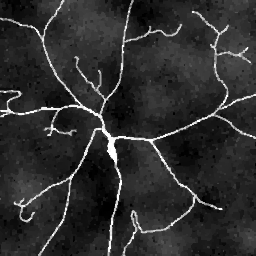} &
\includegraphics[width=\smallreconfigwidth]{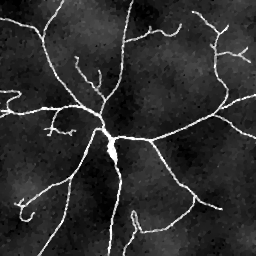}\\
&
\includegraphics[width=\smallreconfigwidth]{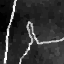} &
\includegraphics[width=\smallreconfigwidth]{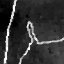} &
\includegraphics[width=\smallreconfigwidth]{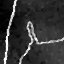}\\[.1cm]
\multirow{1}{*}{\rotatebox{90}{\makebox[\biglabelwidth][c]{\normalsize Random Summation}}}  &
\includegraphics[width=\smallreconfigwidth]{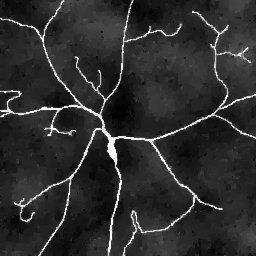} &
\includegraphics[width=\smallreconfigwidth]{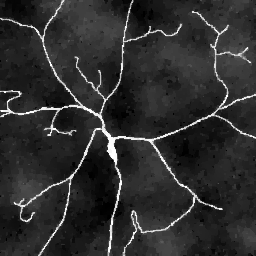} &
\includegraphics[width=\smallreconfigwidth]{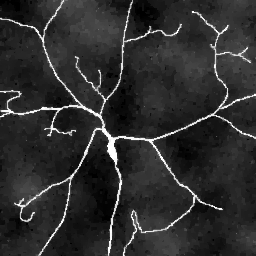}\\
&
\includegraphics[width=\smallreconfigwidth]{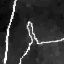} &
\includegraphics[width=\smallreconfigwidth]{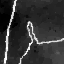} &
 \includegraphics[width=\smallreconfigwidth]{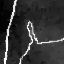}\\[.1cm]
\multirow{1}{*}{\rotatebox{90}{\makebox[\biglabelwidth][c]{\normalsize Subsampling}}}  &
\includegraphics[width=\smallreconfigwidth]{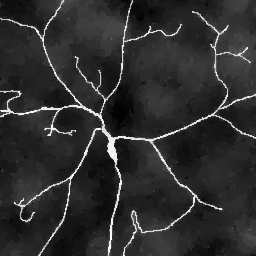} &
\includegraphics[width=\smallreconfigwidth]{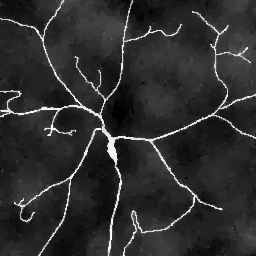} &
\includegraphics[width=\smallreconfigwidth]{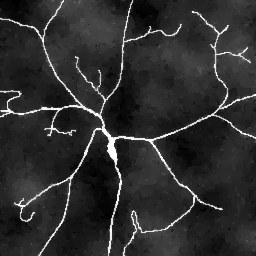}\\
&
\includegraphics[width=\smallreconfigwidth]{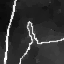}&
\includegraphics[width=\smallreconfigwidth]{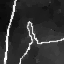} &
\includegraphics[width=\smallreconfigwidth]{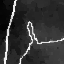}\end{tabular}
\caption{Best performing reconstructions for the different downsampling strategies
(integration downsampling, random summation, and subsampling) and for 
each downsampling strategy, the different mask distributions
(binary spatial, uniform spatial, and uniform phase).
For each set of images, the bottom shows the region indicated by the yellow square in
the ground truth image.  Clearly, significant gains in performance can be achieved by
the compressive architectures over conventional imaging. } 
\label{fig:bigrecon}
\end{figure}

\subsubsection{Mean Subtraction}
\label{sec:meansubtraction}

Generative models for random projection matrices used in CS involve
drawing elements independently from a zero-mean probability
distribution \cite{CS:candes1, hauptnowak, RIP, waheed, JLCS},
and likewise a zero-mean distribution was used to
analyze the coded aperture masks described in Sec.~\ref{sec:cca}.
However, as coded aperture masks with zero mean are not physically
realizable in optical systems, we generate our physically realizable masks from
an appropriately scaled Bernoulli distribution with values $\{0, 1/n\}$. This 
shifting ensures that the coded aperture corresponds to an implementable (\ie nonnegative
and intensity preserving) mask pattern which does satisfies the assumptions needed
to model photon propagation through an optical system.


While necessary to accurately model real-world optical
systems, this shifting negatively impacts the performance of well-established 
$\ell_2$-$\ell_1$ reconstruction algorithms for the following reason.
Several of these algorithms (\eg \cite{sparsa,gpsr}) assume that the $\ell_2$ data fidelity term
$\phi(f) = \tfrac{1}{2} \|Af - y\|_2^2$ (proportional to the negative log Gaussian likelihood) is such that 
$\nabla^2 \phi(x) = A^T A$ can be well-approximated by $\alpha I$, $\alpha > 0$. This assumption is crucial to the performance of these algorithms. 

If we collect measurements using a realizable mask with elements in the range $[0, 1/n]$, the resulting matrix $A$ does not satisfy this condition due to the mean offset. Therefore in the reconstruction we use a mean-shifted sensing matrix
\begin{equation*}
A^0 = A - \mu_A \ones_{m \times n},
\end{equation*}
where  $\mu_A = (1/mn) \ones_{m}^T A \ones_n$ is the mean value of $A$. This new matrix is such that $(A^0)^T A^0 \approx \alpha I$ is a valid assumption. However, to use this matrix in the reconstruction, we need to adjust our estimate accordingly. To compensate for this, we decompose our estimate $f$ into its mean component $\mu_{f}$ and a zero-mean deviation $f^0$: $f = f^0 + \mu_{f}\ones_n$. So then
\begin{equation*}\begin{aligned}
Af &= (A^0 + \mu_{A}\ones_{m \times n})(f^0 + \mu_{f} \ones_{n}) \\
& = A^0 f^0 + n \mu_{A} \mu_{f} \ones_m + \mu_{f} A^0\ones_n + \mu_{A}(\ones_n^T f^0)\ones_m  \\
& \approx A^0 f^0 + n \mu_{A} \mu_{f} \ones_m.
\end{aligned}\end{equation*}
The data $y$ can be decomposed in a similar fashion, so that $y = y^0 + \mu_{y}\ones_m$. Thus a straightforward estimate for the mean of our solution is $\mu_{f} = \mu_y / \mu_{A} n$, and we can use the zero-mean quantities in our data fidelity term
\begin{equation}
\phi^0(f^0) = \tfrac{1}{2} \|A^0 f^0 - y^0\|_2^2
\end{equation}
within the iterative algorithm, which now has the desired property that $\nabla^2 \phi^0(f^0) \approx \alpha I$.

We have motivated this mean subtraction approach from an estimation setting where we first estimate the mean, then deviations about the mean. However, it can be equally motivated from a pure optimization perspective by viewing the mean subtraction step as a preconditioning strategy. The gradient-based reconstruction methods considered are known to exhibit poor convergence when the Hessian has large condition number, and the mean subtraction operation essentially improves the conditioning of the problem by eliminating the single dominant eigenvalue that occurs due to the nonzero mean.

\subsection{Experimental Comparison on Static Images}
\label{sec:numerical}

Here we present numerical experiments supporting the proposed architectures for snapshot compressive imaging. 
We compare the simulated performance of the described imaging architectures in the task of high-fidelity image reconstruction. We utilize a $256 \times 256$ pixel realistic phantom image of a neuron (ground truth image in Fig.~\ref{fig:bigrecon}) which is designed to test the resolution limits of our architectures with its high-resolution features. This image is inspired by a real microscopic image found at \url{http://www.lsi.umich.edu/facultyresearch/labs/bingye/research}. We examine a traditional imaging system and a total of six convolution-based compressive imaging systems.  These systems are constructed by considering each combination of the following design considerations:
\begin{itemize}
\item the mask generation method, including binary spatial (BS) and uniform phase (UP) models,
\item switching among integration downsamping, pseudo-random summation (randomly summing $d/2$ values in each $d_1 \times d_2$ block), and subsampling.
\end{itemize}
To analyze the effect of the number of measurements we collect, we vary the scene-to-sensor downsampling ratio $d$ to be either 4, 16, or 64 (\ie downsampling in both directions by a factor of 2, 4, or 8). While each architecture has its own unique considerations in terms of signal-to-noise efficiency (see Section~\ref{sec:hardware}), we choose to normalize the experiments in such a way as to keep a constant signal-to-noise ratio at the detector.  This is achieved by fixing the variance of the additive white Gaussian noise to be $\sigma^2 = \text{var}{(Af)}/16$ for each architecture when we choose a downsampling ratio of $d=4$. This variance is then fixed for each architecture as we vary the downsampling ratio, the rationale being that this allows us to showcase the impact of the various subsampling operations.

We consider a comparison of many penalization schemes for image reconstruction, such as sparsity in an orthonormal wavelet (Haar) basis, isotropic and anisotropic total-variation \cite{ROF}, as well as an $\ell_1$ sparsity penalty in the overcomplete curvelet basis \cite{candescurvelets}. Curvelets are similar to wavelets in that they capture more spatially localized information than the Fourier components, however they are designed to be more well-adapted for capturing curvilinear singularities in images (\eg edges). To reconstruct the image, we use our own implementation of the SpaRSA algorithm \cite{sparsa} which utilizes the mean subtraction procedure described in Section~\ref{sec:meansubtraction}. A coarse initialization is found by solving an unpenalized least-squares minimization via a conjugate gradient method using only the compressive data. We terminate the reconstruction when the relative change in the iterates falls below a pre-specified tolerance of 1e-3. We choose the regularization weighting to minimize the final reconstruction RMSE, measured by $\text{RMSE}(\widehat{f}) = \|\widehat{f} - \fstar\|_2/\|\fstar\|_2$.  A table of the resulting RMSE values is presented in Table~\ref{tab:RMSE}. 

As evidenced by both the RMSE values and the images, significant gains in performance can be achieved by the compressive architectures. The UP mask architectures slightly outperform the BS-based approaches in this simulation, but the type of subsampling has a greater effect on the quality of the reconstruction. Focusing on the $d=4$ case, we see that subsampling performs best overall, followed by random summation, then integration downsampling. This is readily apparent from the RMSE values and the greater clarity by which we can resolve fine-scale features in the dendrites of the neuron.  This is exactly predicted by the theory: larger coherence yields poorer performance. However, pushing the level of undersampling, we see that simple subsampling is not always the solution, since the performance degrades quite rapidly, as the per-measurement signal to noise ratio does not naturally increase as it does with the other subsampling architectures. An optimal architecture must strike a careful balance between low coherence and robustness to noise.

\begin{table*}[hbt]
\centering \begin{tabular}{@{} cccccccc @{}} 
\toprule
\multicolumn{3}{c}{Sensing Architecture} &	\multicolumn{5}{c}{Reconstruction RMSE (\%)} \\
\cmidrule(r){1-3}  \cmidrule(l){4-8}
Subsampler   & 	Mask Model 	& 	$d$ 	& 	CG Init & 	$\ell_2$-Haar 	& 	$\ell_2$-Aniso. TV & 	$\ell_2$-Iso. TV & $\ell_2$-Curvelet\\
\midrule
Integration  & 	BS 		& 	4 		& 	43.552652 		&	31.908468 	& 	{\bf 23.632185} 	&	23.712212 		&	34.552177	\\
downsampling & 			& 	16 		& 	52.508761 		& 	47.604093 	& 	44.602132 		&	43.339913 		& 	43.606292	\\
             & 			& 	64 		& 	60.435296 		& 	58.592028 	& 	57.537579 		& 	56.960604 		&	56.747754	\\
\cmidrule{2-8}
             & 	US 		& 	4  		& 	43.220564 		&	31.861358	& 	{\bf 23.497272}	& 	23.643347	 	&	34.325474	\\
             & 			& 	16 		& 	52.126070 		&	47.551547	& 	44.400403 		& 	43.069598 		&	43.610772	\\
             & 			& 	64 		& 	59.642746 		&	58.145647	& 	57.118346 		& 	56.684747 		&	56.541953	\\
\cmidrule{2-8}
             & 	UP 		& 	4  		& 	38.474720 		&	30.797415	& 	23.265628 		& 	{\bf 23.214944} 	&	32.475710	\\
             & 			& 	16 		& 	50.284641 		&	46.653984	& 	44.638167 		& 	43.091206 		&	44.231886	\\
             & 			& 	64 		& 	58.815932 		&	58.168464	& 	57.184195 		& 	56.785933 		&	56.558017	\\
\cmidrule{1-8}
Random       & 	BS 		& 	4  		& 	56.890247 		&	34.101763	& 	{\bf 18.992251}	& 	20.535913 		&	35.202771	\\
Summation    & 			& 	16 		& 	62.274874 		& 	51.430633	&	45.573628 		& 	44.644012 		&	45.537800	\\
             & 			& 	64 		& 	65.672366 		&	60.160050 	& 	58.296549 		& 	57.789539 		&	57.630494	\\
\cmidrule{2-8}
             & 	US 		& 	4  		& 	56.544494 		&	33.659520	& 	{\bf 17.484428}	& 	19.021174	 	&	34.322115	\\
             & 			& 	16 		& 	61.950795 		&	50.977833	& 	44.857052 		& 	43.641823 		&	45.217990	\\
             & 			& 	64 		& 	65.350187 		&	59.697311	& 	57.573941 		& 	57.159371 		&	57.096311	\\
\cmidrule{2-8}
             & 	UP 		& 	4  		&	54.473107 		&	29.831022	& 	{\bf 16.033758}	& 	17.728624 		&	33.651808	\\
             & 			& 	16 		& 	60.297045 		& 	48.902002	&	44.939914		& 	43.733698 		&	45.219903	\\
             & 			& 	64 		& 	64.317781 		& 	60.909768	&	57.360741		& 	57.106131 		&	56.933892	\\
\cmidrule{1-8}
Pointwise    & 	BS 		& 	4 		& 	61.578677		&	39.742969 	& 	{\bf 17.129559}	& 	18.611897		& 	35.793504 	\\
Subsampling  & 			& 	16 		& 	67.787137		& 	65.569753 	& 	51.912236		&  	50.842530		& 	52.625216 	\\
             & 			& 	64 		& 	69.244807		&	69.202225 	&	64.772033		& 	64.693113		& 	64.741188 	\\				
\cmidrule{2-8}
             & 	US 		& 	4  		& 	61.578677 		&	39.780250	& 	{\bf 17.200487}	& 	18.703763	 	&	35.813267	\\
             & 			& 	16 		& 	67.787137 		&	65.570111	& 	51.928327 		& 	50.883713 		&	52.629357	\\
             & 			& 	64 		& 	69.244807 		&	69.202225	& 	64.791854 		& 	64.687147 		&	64.741653	\\
\cmidrule{2-8}
             & 	UP 		& 	4 		& 	61.005834		&	36.260055 	& 	{\bf 14.108310}	& 	15.883293		& 	34.282494	\\
             & 			& 	16 		& 	67.675516		& 	64.606384 	& 	48.280444		&  	47.379639		& 	48.924798	\\
             & 			& 	64 		& 	69.187676		& 	69.189077 	&	63.591662		& 	63.514672		& 	63.350203	\\				
\midrule
\multicolumn{2}{c}{Conventional Imager}
              			&	4  		& 	38.157537 		&	36.654837	& 	36.067380 		& 	{\bf 32.946067}	&	33.157436	\\
             &			&	16 		& 	50.137698 		&	50.117667	& 	49.836071 		& 	47.932575 		&	45.466410	\\
             &			&	64 		& 	58.873484 		&	58.873455	& 	58.472819 		& 	57.806412 		&	56.764936	\\
\bottomrule
\end{tabular}
\caption{Summary of the RMSE values for the different mask generation and downsampling schemes, including the conjugate gradient initialization, and the various reconstruction methods considered. The images associated with the best performing methods indicated by the bold RMSE values are shown in Fig.~\ref{fig:bigrecon}}
\label{tab:RMSE}
\vspace{-0.25in}
\end{table*}

\section{Coded Aperture Keyed Exposure Sensing for Dynamic Scenes}
\label{sec:video}

Here we detail our compressive video acquisition method that combines coded apertures and keyed exposures to address all the competing challenges detailed in Sec.~\ref{sec:probform}.
Specifically, in our CAKE imaging method, each observed frame $y_k$ is given by an exposure of $B$ high-rate coded observations:
\begin{equation}
y_k = \sum_{t=1}^B A_{(k-1)B + t} \fstar_{(k-1)B + t} + w_k,
\end{equation}
where each $A_t$ describes an AMM CCA sensing matrix. 
Note that since in our theory, the downsampling operator $\downsamp_\text{sub}$ is a structured nonrandom operator, we can rewrite the above as
\begin{equation}
y_k = \downsamp_\text{sub} \left[\sum_{t=1}^B R_{(k-1)B + t} \fstar_{(k-1)B + t}\right] + w_k.
\end{equation}
Hence all that is required to implement this sensing paradigm is modulating the coded aperture mask over the $B$-frame exposure time. Because of this, one can think of our system as performing a coded exposure acquisition \emph{for each coded aperture element}. 

Since our sensing strategy is independent across each low-resolution frame, and also for simplicity of presentation, we only consider the recovery of a length $B$ block of frames from a single snapshot image in our theoretical analysis. 

\begin{theorem}[Coded Aperture Keyed Exposure Sensing]
\label{thm:CAKE} Let $A = [A_1 \cdots A_B]$ be an $m \times nB$ sensing matrix for the CAKE system where the coded aperture pattern for each $A_t$ is drawn i.i.d.\ from a suitable probability distribution. Then there exists constants $c_1, c_2 \ge 0$ depending on $\delta_{s}$ such that for any
\begin{equation}
m \ge c_1 s^2 \log(nB),
\end{equation}
$A$ is $\RIP(s,\delta_s)$ with probability exceeding
\begin{equation}
	1 - 2n^2B^2 e^{-c_2 m/s^2}.
\end{equation}
\end{theorem}
Note here that $s$ denotes the sparsity of the first $B$ frames of the video sequence, rather than simply the sparsity of an individual frame as in Thm.~\ref{thm:BCCB}. The proof of Thm.~\ref{thm:CAKE} is presented in Appendix~\ref{sec:CAKEProof}. Similar to Thm.~\ref{thm:BCCB}, this proof assumes that the entries of each $h_t$, $t = 1,\ldots,B$ are generated i.i.d.\ according to the scaled Rademacher distribution \eqref{eq:rademacher}. Again, it is straightforward to extend the result to other mask generating distributions.

\subsection{Sparse Transformation Sensing}

It is most often the case that the original frames $\fstar$ are not sparse, but can be sparsely represented by a temporal transform of the frames. For simplicity, consider the first coded measurement. Within this acquisition, the coefficient sequence
\begin{equation*}
\thetastar_k = \sum_{t=1}^B W_{k,t} \fstar_t, \quad k = 1, \ldots, B,
\end{equation*}
may be more sparse for a well-chosen sparse temporal transform $W$. Notationally this is equivalent to $\thetastar = (W \otimes I_n) \fstar$, where $\otimes$ denotes the Kronecker matrix product. A preferred sensing strategy would be to use our RIP-satisfying CAKE acquisition to sense the coefficient sequence $\thetastar$:
\begin{equation}
y = \sum_{k=1}^B A_k \thetastar_k + w. \label{eq:cake1}
\end{equation}
Surprisingly, this can be accomplished using an identical architecture, with some slight adjustment to the coded aperture mask patterns used during the keyed acquisition. 

To see this, we examine the resulting sensing matrix in terms of $\fstar$:
\begin{equation*}
\begin{aligned}
\sum_{k=1}^B A_k \thetastar_k 
&= \sum_{k=1}^B A_k \sum_{t=1}^B W_{k,t} \fstar_t
= \sum_{t=1}^B \left[ \sum_{k=1}^B W_{k,t} A_k \right] \fstar_t \\
&= \sum_{t=1}^B A^W_t \fstar_t,
\end{aligned}
\end{equation*}
where we define
\begin{equation}
A^W_t = \sum_{k=1}^B W_{k,t} A_k.
\end{equation}
Therefore \eqref{eq:cake1} is also a CAKE acquisition using the sensing matrices $A^W_t$. Because of the linear dependence on the generating mask patterns, this amounts to using \emph{the an identical architecture with adjusted mask patterns}
\begin{equation}
h^W_t = \sum_{k=1}^B W_{k,t} h_k.
\end{equation}
If we denote $h^W = (h^W_1,\ldots,h^W_B)$, then this can be written more simply as $h^W = (W^T \otimes I_n) h = (W \otimes I_n)^T h$. Hence the CAKE system can adapt to any temporal sparse coding of the video sequence over the block of coded frames. In summary, to incorporate the sparse transform $W$ applied to the frames, we simply apply the adjoint transform to the independently generated mask patterns. 

A useful transformation that allows the exploitation of dependencies between frames is to assume that the \emph{difference} between subsequent frames are sparse. In this case, we select $W = \grad$, as defined in \eqref{eq:grad}. In particular, we sense using the coded sequence of aperture patterns $h^\grad$ where
\begin{equation*}
h^\grad_t = \begin{cases} h_k - h_{k+1} & k = 1, \ldots, B-1, \\ 
	h_B & k = B, \end{cases}
\end{equation*}
and the generating aperture patterns $h$ are drawn from a suitable distribution.

It is of interest to note that since the CAKE sensing matrix is a concatenation of downsampled Toeplitz matricies, this sensing strategy has clear connections to \cite{WaheedMud} where they consider concatenations of Toeplitz matricies as a sensing matrix for performing multiuser detection in wireless sensor networks. The important conceptual link is that their sensing matrix is used to determine a sparse set of simultaneously active users, where in our system, we are using it to infer a sequence of sparse frames, or a sequence of sparse difference frames.

\subsection{Implementation and Normalization Details}

Recall that these RIP-satisfying sensing matrices are generated using a zero-mean probability distribution and cannot be directly implemented in an optical system. As before, we apply an affine transformation to each of the $A_k$ so that the resulting mask elements for each $h_k$ are within the range $[0, 1/n]$. If we are generating a sensing matrix that satisfies RIP with respect to the difference frames, then even a binary generating distribution will cause the resulting mask patterns to have elements that are neither fully open nor fully closed. This can be accomplished using half-toning as discussed in Sec.~\ref{sec:hardware}. If such an implementation is difficult, then one strategy would be to set these intermediate values to $0$ or $1/n$ at random with equal probability. 

\subsection{Reconstruction for Video}

Given measurements from the proposed CAKE imaging system, where we have designed the mask patterns for sparsity in a temporal transform, we recover the frames by solving
\begin{equation*}
\begin{aligned}
\widehat{\theta} &= \argmin_\theta \tfrac{1}{2} \|A \theta - y\| + \tau \|\theta\|_1 \\
\widehat{f} &= (W^{-1} \otimes I_n) \widehat{\theta}.
\end{aligned}
\end{equation*}
For the case of sparse difference frames ($W = \grad$), improvements in empirical performance are made by penalizing the total variation of the first frame, instead of simply the sparsity in that frame:
\begin{equation*}
\begin{aligned}
\widehat{\theta} &= \argmin_\theta \tfrac{1}{2} \|A \theta - y\| + \tau_\text{TV}\|\theta_1\|_\text{TV} 
	+ \tau_1 \sum_{k=2}^B \|\theta_k\|_1\\
\widehat{f} &= (L \otimes I_n) \widehat{\theta},
\end{aligned}
\end{equation*}
here $L = \grad^{-1}$ is a lower triangular matrix of all ones.

In addition, by using more than one low-rate observation per reconstruction step, we are typically able to improve performance by coding the difference frames across more than one length-$B$ block of high-rate frames. This is the reconstruction technique we use in the experimental results section. In previous work, we noticed that when the amount of processing time allotted per frame is held constant, the accuracy generally increases with the number of frames processed simultaneously.  However, one simply cannot solve for arbitrarily many frames simultaneously, as the improvement in accuracy diminishes when the size of the problem is such that only a very small number of reconstruction iterations can be run within the allotted time.  We refer the reader to \cite{MarciaEUSIPCO} for details.

For video sequences of longer duration, we may wish to process the video in an online fashion, solving for only a few blocks of frames simultaneously. In many applications, such as surveillance and monitoring, the video frames are strongly correlated, and the solution to a previous block of frames may be used as an initialization to the algorithm to solve the next block of frames. This estimate will generally be very accurate, and therefore, relatively few iterations are needed to obtain a solution to each optimization problem.

\subsection{Numerical Experiments}

\begin{figure}[t]
\begin{center}
\includegraphics[width=\columnwidth]{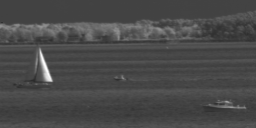}
\caption{Example frame (frame 21) from the ground truth video sequence used in the numerical experiments.}
\label{fig:singleframe} 
\end{center}
\end{figure} 

In this section, we demonstrate the effectiveness of the proposed CAKE architecture at successfully recovering a video sequence of short-wave infrared (SWIR) data collected (courtesy of Jon Nichols at NRL) by a short-wave IR ($0.9-1.7\mu\text{m}$) camera. The camera is based on a $1024\times 1280$ InGaAs (Indium, gallium
arsenide) focal plane array with $20\mu\text{m}$ pixel pitch.  Optically, the camera was built around a fixed, $f/2$ aperture and provides a $6^{\circ}$ field of view along the diagonal with a focus range of $50\text{m}\rightarrow\infty$.  Imagery were output at the standard 30Hz frame rate with a 14 bit dynamic range. An example frame is shown in Fig.~\ref{fig:singleframe}. In this sequence, the three boats are traveling at different velocities with respect to the slowly-changing background of the waves. The size of each frame is $128 \times 256$, and we consider reconstructing 28 frames of the sequence. 

We consider CAKE observations where we downsample spatially by a factor of 2 in both directions ($d_1 = d_2 = 2$), and use a coded exposure block length of $B = 4$. We compare acquiring the sequence with independent mask codes, and mask codes designed to exploit the sparse difference frames directly. We reconstruct the entire video sequence using all 7 low-resolution low-rate frames using a total variation penalty on the first frame, and $\ell_1$ sparsity penalty on all subsequent difference frames. We optimize the regularization parameters to minimize the reconstruction error. For comparison, we consider traditionally captured data (\ie by simply averaging over $d_1 \times d_2 \times B$ blocks of the spatiotemporal video sequence). To interpolate this data to the original resolution of the video sequence, we consider both nearest-neighbor and spline interpolation. 

\begin{figure}[t]
\begin{center}
\includegraphics[width=\columnwidth]{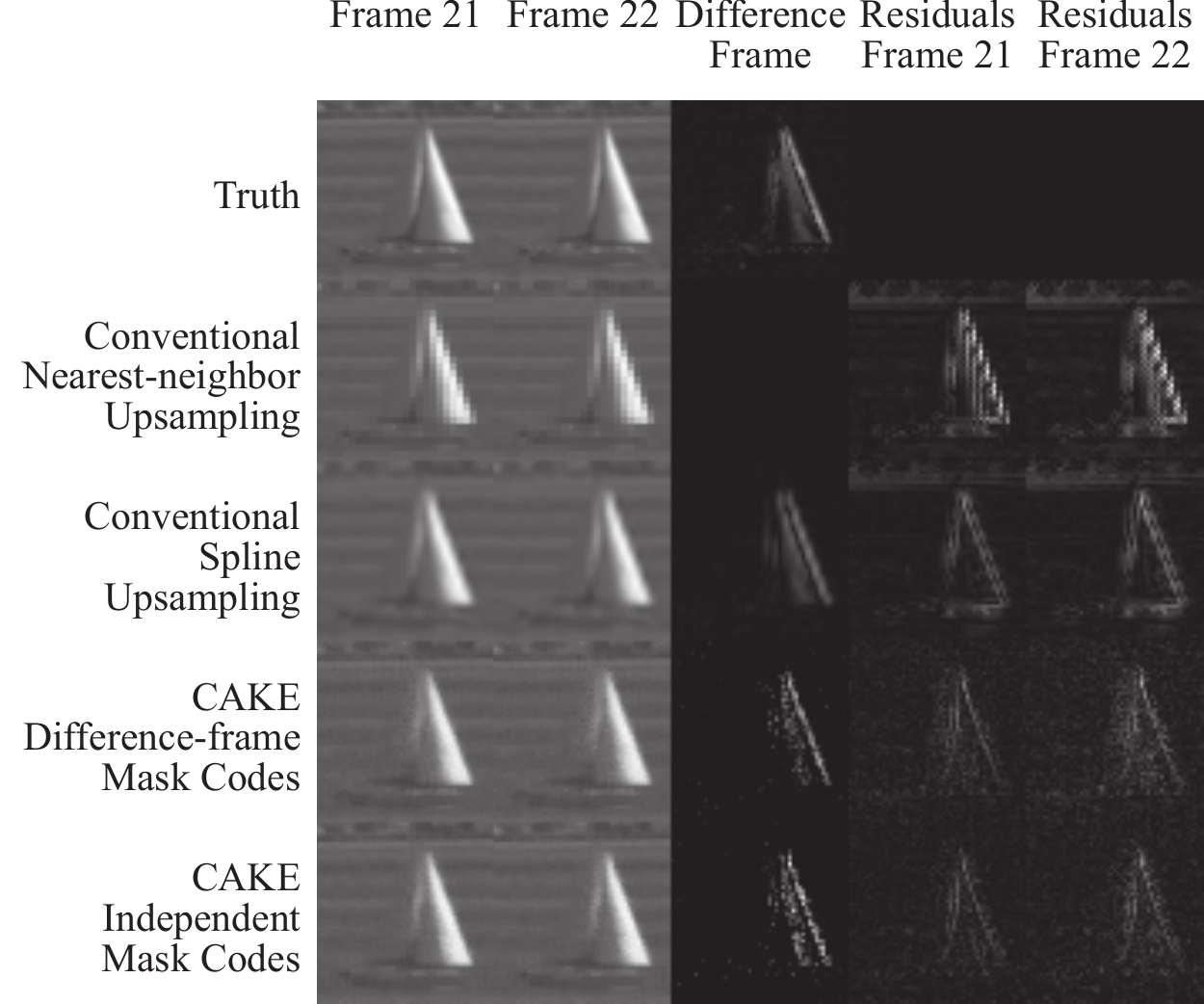}
\caption{Results obtained for the sailboat ROI for an example pair of frames (frames 21 and 22). Shown are the true frames, the reconstruction from traditionally sampled data, and the reconstruction from CAKE sensed data. For comparison we show the residuals as compared with the truth, as well as the difference between the frames. Note that using nearest neighbor interpolation results in no motion over the block of frame, hence a zero difference frame.}
\label{fig:SailboatPanel} 
\end{center}
\end{figure}

We show the estimates for the different acquisition and reconstruction methods in Fig.~\ref{fig:SailboatPanel}. Here we focus only on a ROI of the sailboat. We see that the CAKE sensing is able to reconstruct the scene with a higher spatial and temporal resolution. This is evidenced in the residuals, which include much less critical scene structure than the conventional system. We see from the examination of the difference frame that the nearest-neighbor reconstruction from conventionally sampled data yields no motion over the two frames and suffers from poor spatial resolution. Using spline interpolation helps improve the spatial resolution, but it is still insufficient to recover the scene with high temporal resolution, as can be seen in the blur on the leading edge of the sail. Numerically we quantify the performance over the entire video sequence in terms of the RMSE (\%), $100 \cdot \|\widehat{f} - \fstar\|_2/\|\fstar\|_2$, calculated both over the entire frame, and only over the sailboat ROI. This is tabulated in Table~\ref{tab:CAKERMSE}. In summary we see that the CAKE acquisitions are able to outperform traditionally sampled video in terms of reconstruction accuracy and reconstructing salient motion.

It should be noted that there are limitations to the CAKE system in the presence of strong motion. In this case, the sparsity level of the difference frames may drastically increase as the previous frame ceases to be a good prediction of the next frame. As such, the RIP bound in Theorem~\ref{thm:CAKE} states the number of measurements we require to reconstruct the scene must necessarily increase, requiring either a faster temporal resolution measurements or higher resolution FPA to achieve the same accuracy. Because of this balance, a system designer may need to make important engineering tradeoffs to implement CAKE acquisition for a particular application.

%

\begin{table}[t]
\centering
\begin{tabular}{@{} lcc @{}}   
\toprule
                            & \multicolumn{2}{c}{Reconstruction RMSE (\%)} \\
\cmidrule(l){2-3}
Sensing Architecture        & Full Frame & Sailboat ROI \\
\midrule
Conventional (Nearest-neighbor)    & 5.5163       (5.5305) &                10.3525            (10.4241)  \\ 
Conventional (Spline)              & 3.7654       (3.8047) &      \phantom{1}5.9335  (\phantom{1}6.0377)  \\
CAKE (Difference-frame Codes)      & 3.5840       (3.9183) &      \phantom{1}5.6971  (\phantom{1}7.0421)  \\
CAKE (Independent Codes)           & {\bf 2.9079} (3.0932) & {\bf \phantom{1}4.3266} (\phantom{1}4.9604)  \\
\bottomrule
\end{tabular}
\caption{Reconstruction RMSE achieved for the conventional and CAKE architectures over the video sequence. Results are reported both for the full frame and the ROI of the sailboat. Due to boundary issues, we report the RMSE discounting the first and last block of $B$ frames. The RMSE values for the entire sequence are given in parentheses.}
\label{tab:CAKERMSE}
\end{table}

\section{Conclusions}
\label{sec:conc}

Compressed sensing offers a strong theoretical foundation for sparse signal recovery.
However, practical and implementable imaging system designs based on CS theory
have lagged far behind.  
In this paper, we demonstrate how CS principles can be
applied to physically realizable optical system designs, namely coded aperture imaging.
Numerical experiments show that CS methods can help resolve high resolution
features in images and videos that conventional imaging systems cannot.
We have also demonstrated that our CAKE acquisition system can recover video sequences from highly under-sampled data in much broader settings than those considered in initial coded exposure studies. However, our derived theoretical limits show that there are important tradeoffs involved that depend on the spatial sparsity, and temporal similarity of the scene that ultimately govern the accuracy of our reconstructions for a specified number of compressive measurements.
Finally, we note that our proposed approach performs well 
in high signal-to-noise ratio settings, but like all CS reconstructions, 
it can exhibit significant artifacts in photon-limited settings.


Two practical issues associated with coded aperture imaging in general are the blur due to misalignment of the mask and diffraction and interference effects. Noncompressive aperture codes have been developed to be robust to these effects \cite{diffractionEffects}. One important avenue for future research is the development of {\em compressive} coded aperture makes with similar robustness properties. Noncompressive coded apertures have also been shown useful in inferring the depth of different objects in a scene; similar inference may be possible with the compressive coded apertures described in this paper.

\section*{Acknowledgements}
The authors would like to thank Prof. Justin Romberg and Dr. Kerkil Choi for several valuable
discussions.

\appendix

\subsection{Proof of the RIP for Spatial-Domain CCA Sensing}
\label{sec:proof}
Here we present the proof that appropriately scaled compressive coded apertures satisfy the RIP.


\begin{IEEEproof}[Proof of Theorem~\ref{thm:BCCB}]
In the interest of notational simplicity, it is easier to work on the two-dimensional images versus their one-dimensional vectorial representations. For concreteness, we assume the entries of $h$ are generated i.i.d.\ according to the scaled Rademacher distribution 
\begin{equation*}
h_{k_1,k_2} = \begin{cases} \phantom{-}\sqrt{d/n} & \text{ with probability } 1/2, \\
									 -\sqrt{d/n} & \text{ with probability } 1/2, \end{cases}
\end{equation*}
for $k_1 = 1, \ldots, n_1$ and $k_2 = 1, \ldots, n_2$. The proof uses the same techniques as that of Theorem 4 in \cite{haupttoeplitz}, where the RIP is established by shifting the analysis of the submatrices of $A$ to the entries of the Gram matrix $G = A^T\!A$ by invoking Ger\v{s}gorin's disc theorem \cite{Verga}. This theorem states that the eigenvalues of an $n \times n$ complex matrix $G$ all lie in the union of $n$ discs $d_j(c_j,r_j)$, $j = 1,2,\ldots,n$, centered at $c_j = G_{j,j}$ with radius
\begin{equation*}
r_j = \sum_{i=1 \atop i \ne j}^m |G_{i,j}|.
\end{equation*}
In essence, we show that with high probability $G \approx I$, so that the eigenvalues of $G$ are clustered around one with suitably high probability. 

From the discussion above, we have that
\begin{equation*}\begin{aligned}
A_{l,k} = h_{\text{mod}[(l_1-1)d_1 - k_1 + 1,n_1]+1, \text{mod}[ (l_2-1)d_2 - k_2 + 1,n_2]+1 },
\end{aligned}\end{equation*}
so the entries of the resulting Gram matrix $G$ are
\begin{equation}\begin{aligned}\label{eq:grammatrix}
G_{p,q} \ &\!\! = \sum_{l_1 = 1}^{n_1/d_1}\sum_{l_2 = 1}^{n_2/d_2} \\
&h_{\text{mod}[(l_1-1)d_1 - p_1 + 1,n_1]+1, \text{mod}[ (l_2-1)d_2 - p_2 + 1,n_2]+1 } \cdot \\
&h_{\text{mod}[(l_1-1)d_1 - q_1 + 1,n_1]+1, \text{mod}[ (l_2-1)d_2 - q_2 + 1,n_2]+1 }.
\end{aligned}\end{equation}
From the normalization introduced for $h$, we can show that $\expect [G] = I$.  Now we need to bound the deviation about the mean via concentration. We consider first the diagonal entries of $G$, when $p = q$. Each term is a sum of $n = n_1 n_2$ bounded i.i.d.\ entires, and applying Hoeffding's inequality yields
\begin{equation*}
\prob(|G_{q,q} - 1| \ge \delta_{\text{d}}) \le 2 \exp\left(\frac{-2n\delta_{\text{d}}^2}{d} \right).
\end{equation*}
Next we consider the off-diagonal entries in the special case that either $\text{mod}(p_1-q_1,d_1) \ne 0$ or $\text{mod}(p_2-q_2,d_2) = 0$. In this case, each of the terms in the summand in \eqref{eq:grammatrix} picks out a \emph{different} set of coefficients from $h$, and hence there are no dependencies between different terms in the sum, hence it is also a sum of $n = n_1 n_2$ bounded i.i.d.\ entries, and thus from Hoeffding's inequality, we have
\begin{equation*}
\prob(|G_{p,q}| \ge \delta_{\text{o}}/s) \le 2 \exp \left( \frac{-n\delta_\text{o}^2}{2ds^2} \right).
\end{equation*}
Lastly, we have the case that both $\text{mod}(p_1-q_1,d_1) = 0$ and $\text{mod}(p_2-q_2,d_2) = 0$ with $p \ne q$. This case deserves special care since each of the terms in the summand in \eqref{eq:grammatrix} picks out \emph{the same} set of coefficients from $h$. Therefore there necessarily exist dependencies within the terms of the sum. However, due to the special nature by which we select the coefficients, we can partition the sum $G_{p,q}$ into two sums, denoted $S_1$ and $S_2$ such that each of these is a sum of $n/2d$ independent terms. We can then apply the Hoeffding bound to each of these to yield
\begin{equation*}
\prob(|S_i| \ge \delta_{\text{o}}/2s) \le 2 \exp \left( \frac{-n\delta_\text{o}^2}{4 ds^2} \right), i = 1,2.
\end{equation*}
Then applying the union bound gives us
\begin{equation*}
\prob(|G_{p,q}| \ge \delta_{\text{o}}/s) \le 4 \exp \left( \frac{-n\delta_\text{o}^2}{4 ds^2} \right).
\end{equation*}
Since this latter bound decays more slowly, and in the interest of simplicity, we overbound using this latter expression. 
What remains is to now apply the union bound over all the diagonal and off-diagonal elements. For this we need to count the number of elements in the union. For the diagonal entries, there are clearly $n$ elements. And since the entries of $G$ are symmetric, we only have $n(n-1)/2$ remaining terms. Hence taking $\delta_\text{d} = \delta_\text{o} = \delta_s/2$,
\begin{equation*}\begin{aligned}
&\prob(A \text{ does not satisfy } \RIP(s,\delta_s)) \\
& \quad \le 2n \exp\left(\frac{-n\delta_s^2}{2d} \right) + 4[n(n-1)/2] \exp \left( \frac{-n\delta_s^2}{16 ds^2} \right) \\
& \quad \le [2n + 2n(n-1)] \exp \left( \frac{-n\delta_s^2}{16 ds^2} \right) \\
& \quad = 2n^2 \exp \left( \frac{-n\delta_s^2}{16 ds^2} \right) \le 2n^2 \exp \left( \frac{-c_2 n}{ds^2} \right),
\end{aligned}\end{equation*}
where $c_2 \le \delta_s^2/16$. If $n \ge 2$, this probability is less than one provided $n/d \ge c_1 s^2 \log(n)$ where $c_1 \ge 3/c_2$, noting that $m=n/d$ establishes the theorem.
\end{IEEEproof}

\subsection{Proof of RIP for CAKE}
\label{sec:CAKEProof}

\begin{IEEEproof}[Proof of Theorem~\ref{thm:CAKE}]
This section details the proof of Theorem~\ref{thm:CAKE}. Our strategy here is the same as that of a single frame, we bound the entries of the Gram matrix. First note that in this case the Gram matrix has a certain block structure; since $A = [A_1 \cdots A_B]$, we have
\begin{equation*}
G = A^TA = 
\begin{bmatrix}
	A_1^T A_1    & A_1^T A_2    & \cdots    & A_1^T A_B  \\
	A_2^T A_1    & A_2^T A_2    & \cdots    & A_2^T A_B  \\
	\vdots       & \vdots       & \ddots    & \vdots     \\
	A_B^T A_1    & A_B^T A_2    & \cdots    & A_B^T A_B
\end{bmatrix}.
\end{equation*}
This block structure allows us to utilize the results established in Appendix~\ref{sec:proof}. For the diagonal elements of $G$, we simply use the same bound for one particular block $A_k^T A_k$, since this is the situation for a sensing a single frame. Therefore
\begin{equation*}
\prob(|G_{q,q} - 1| \ge \delta_{\text{d}}) \le 2 \exp\left(\frac{-2n\delta_{\text{d}}^2}{d} \right).
\end{equation*}
For the off-diagonal entries of $G$, we need to consider the off-diagonal entries of $A_k^T A_k$ and any entry of $A_k^T A_l$, $k \ne l$. Recall that in the Appendix~\ref{sec:proof}, we simply used the worst-case slower-decaying bound when the off-diagonal entry of the Gram matrix was no longer a sum of all independent terms. Here we use the same overbound and hence have
\begin{equation*}
\prob(|G_{p,q}| \ge \delta_{\text{o}}/s) \le 2 \exp \left( \frac{-n\delta_\text{o}^2}{2ds^2} \right).
\end{equation*}
Lastly, we need to perform a union bound over all the entries of the matrix. We have $nB$ diagonal entries, and exploiting symmetry, we only have $nB(nB-1)/2$ remaining entries. Hence taking $\delta_\text{d} = \delta_\text{o} = \delta_s/2$,
\begin{equation*}\begin{aligned}
&\prob(A \text{ does not satisfy } \text{RIP}(s,\delta_s)) \\
& \quad \le 2nB \exp\left(\frac{-n\delta_s^2}{2d} \right) + 4[nB(nB-1)/2] \exp \left( \frac{-n\delta_s^2}{16 ds^2} \right) \\
& \quad \le [2nB + 2nB(nB-1)] \exp \left( \frac{-n\delta_s^2}{16 ds^2} \right) \\
& \quad = 2n^2B^2 \exp \left( \frac{-n\delta_s^2}{16 ds^2} \right) \le 2n^2B^2 \exp \left( \frac{-c_2 n}{ds^2} \right),
\end{aligned}\end{equation*}
where $c_2 \le \delta_s^2/16$. If $nB \ge 2$, this probability is less than one provided $n/d \ge c_1 s^2 \log(nB)$ where $c_1 \ge 3/c_2$, again noting that $m=n/d$ establishes the theorem.
\end{IEEEproof}

\ifCLASSOPTIONcaptionsoff
  \newpage
\fi

\bibliographystyle{IEEEbib}
\bibliography{TIPCoded}

\begin{IEEEbiography}{Zachary T.~Harmany}
Zachary Harmany received the B.S. (magna cum lade, with honors) in Electrical Engineering and B.S. (cum lade) in Physics from The Pennsylvania State University in 2006. Currently, he is a Ph.D. student in the department of Electrical and Computer Engineering at Duke University. In 2010 he was a visiting researcher at The University of California, Merced. His research interests include nonlinear optimization, functional neuroimaging, and signal and image processing with applications in medical imaging, astronomy, and night vision. 
\end{IEEEbiography}

\begin{IEEEbiography}{Roummel F.~Marcia}
Roummel Marcia received his B.A. in Mathematics from Columbia  University in 1995 and his Ph.D. in
Mathematics from the University of California, San Diego in 2002.  He was a Computational and Informatics in Biology and Medicine Postdoctoral Fellow in the Biochemistry Department at the University of Wisconsin-Madison and a Research Scientist in the Electrical and Computer Engineering at Duke University.  He is currently an Assistant Professor of Applied Mathematics at the University of California, Merced.  His research interests include nonlinear optimization, numerical linear algebra, signal and image processing, and computational biology.
\end{IEEEbiography}


\begin{IEEEbiography}{Rebecca M.~Willett} (SM '02, M '05, SM '11) is an assistant professor in the Electrical and Computer Engineering Department at Duke University. She completed her PhD in Electrical and Computer Engineering at Rice University in 2005. Prof. Willett received the National Science Foundation CAREER Award in 2007, is a member of the DARPA Computer Science Study Group, and received an Air Force Office of Scientific Research Young Investigator Program award in 2010. Prof. Willett has also held visiting researcher positions at the Institute for Pure and Applied Mathematics at UCLA in 2004, the University of Wisconsin-Madison 2003-2005, the French National Institute for Research in Computer Science and Control (INRIA) in 2003, and the Applied Science Research and Development Laboratory at GE Healthcare in 2002. Her research interests include network and imaging science with applications in medical imaging, wireless sensor networks, astronomy, and social networks. 
\end{IEEEbiography}


\vfill


\end{document}